\documentclass[]{emulateapj}

\def\bibfiles{biblio}
\def\aareferences{\bibliographystyle{apj}
                  \bibliography{aajour,\bibfiles}}

\def\rmit#1{{\it #1}}              
\def\specchar#1{{\sc #1}}
\def\FeI{\mbox{Fe\,\specchar{i}}}

\def\NaIDone{\mbox{Na\,\specchar{i}\,\,D$_1$}}
\def\NaIDtwo{\mbox{Na\,\specchar{i}\,\,D$_2$}}
\def\SiI{\mbox{Si\,\specchar{i}}}
\def\HeI{\mbox{He\,\specchar{i}}}
\def\CaIIH{\mbox{Ca\,\specchar{ii}\,\,H}}       

\def\ie{\rmit{i.e.}}
\def\eg{\rmit{e.g.}}

\usepackage{natbib,graphicx}

\usepackage{color}

\shorttitle{Magneto-acoustic energy in umbral atmosphere}

\shortauthors{Felipe et al.}

\begin{document}

\title{Magneto-acoustic wave energy from numerical simulations of an observed sunspot umbra}

\author{T. Felipe\altaffilmark{1,2}, E. Khomenko\altaffilmark{1,2,3} and M. Collados\altaffilmark{1,2}}
\email{tobias@iac.es}

\altaffiltext{1}{Instituto de Astrof\'{\i}sica de Canarias, 38205,
C/ V\'{\i}a L{\'a}ctea, s/n, La Laguna, Tenerife, Spain}
\altaffiltext{2}{Departamento de Astrof\'{\i}sica, Universidad de La Laguna, 38205, La Laguna, Tenerife, Spain} 
\altaffiltext{3}{Main Astronomical Observatory, NAS, 03680, Kyiv,
Ukraine}

\begin{abstract}
We aim at reproducing the height dependence of sunspot wave signatures obtained from spectropolarimetric observations through 3D MHD numerical
simulations. A magneto-static sunspot model based on
the properties of the observed sunspot is constructed and perturbed at the photosphere introducing the fluctuations measured with the \SiI\ $\lambda$ 10827 \AA\ line. The results of the simulations are
compared with the oscillations observed simultaneously at different heights from
the \HeI\ $\lambda$ 10830 \AA\ line, the \CaIIH\ core and the \FeI\ blends in the wings
of the \CaIIH\ line. The simulations show a remarkable agreement with the observations. They reproduce the velocity maps and power spectra at the formation heights of the observed lines, as well as the phase and amplification spectra between several pair of lines. We find that the stronger shocks at the chromosphere are accompanied  with a delay between the observed signal and the simulated one at the corresponding height, indicating that shocks shift the formation height of the chromospheric lines to higher layers. Since the simulated wave propagation matches very well the properties of the observed one, we are able to use the numerical calculations to quantify the energy contribution of the magneto-acoustic waves to the chromospheric heating in sunspots. Our findings indicate that the energy supplied by these waves is too low to balance the chromospheric radiative losses. The energy contained at the formation height of the lowermost \SiI\ $\lambda$ 10827 \AA\ line in the form of slow magneto-acoustic waves is already insufficient to heat the higher layers, and the acoustic energy which reaches the chromosphere is around 3-9 times lower than the required amount of energy. The contribution of the magnetic energy is even lower.

\end{abstract}



\section{Introduction}

The question about the processes which heat the stellar outer atmospheres is one of the most intriguing unanswered problems in astrophysics. Among all the mechanisms which have been proposed to account for these energy losses, two of them seem to be the most promising ones: mechanical heating by upward-propagating waves generated in the convection zone \citep{Alfven1947,Biermann1948,Schwarzschild1948} and Joule heating driven by magnetic field reconnection and resistive dissipation of electric currents \citep{Parker1983, Heyvaerts+Priest1983}. The work by \citet{SocasNavarro2005} concluded that likely the Joule heating mechanism cannot provide the dominant source to heat the sunspot chromosphere, leaving the energy transport by waves as a plausible candidate.

The magnitude of the energy supplied to the quiet Solar chromosphere by high-frequency acoustic waves has been widely studied over the last years. \citet{Fossum+Carlsson2005, Fossum+Carlsson2006} analyzed temporal series of the quiet Sun observations obtained with the Transition Region And Coronal Explorer (TRACE) in two continuum bands at 1600 and 1700 \AA. From the comparison of these observations with one-dimensional (1D) numerical simulations, they retrieved an acoustic flux supplied to the chromosphere by high frequency waves (at 5--28 mHz) of at least 10 times lower than the required amount of energy to account for the chromospheric radiative losses. However, some authors have criticized this result. Using three-dimensional (3D) numerical simulations, \citet{WedemeyerBohm+etal2007} and \citet{Cuntz+etal2007} have argued that the limited spatial resolution of TRACE, around 1'', hides a factor of 10 in the short-period energy flux. \citet{Kalkofen2007} shared this point of view, and claimed that the observations of chromospheric radiation support the heating by the dissipation of acoustic waves. \citet{Carlsson+etal2007} analyzed high-resolution time sequences of \CaIIH\ filtergrams from SOT/HINODE, finding a larger acoustic power, but still too low to balance the chromospheric radiative losses. An even larger value was found by \citet{BelloGonzalez+etal2009}. Recently, using the IMAX spectropolarimeter \citep{MartinezPillet+etal2010} onboard SUNRISE \citep{Barthol+etal2010, Solanki+etal2010}, \citet{BelloGonzalez+etal2010} obtained an acoustic flux at a height of 250 km above the photosphere only around 2 times lower than the chromospheric radiative losses. However, \citet{Fleck+etal2010} argue that energy fluxes derived from measured high frequency waves may be largely overestimated, since their power may be caused by line formation effects in a dynamic atmosphere rather than propagating high frequency acoustic
waves.

With regards to quiet Sun waves with frequencies in the 3 and 5 mHz bands, according to \citet{Beck+etal2009}, their wave energy is insufficient to maintain the temperature increase at layers higher than 500 km above the photosphere.

Some works have pointed out that the energy flux associated to other wave modes may be added up to balance the chromospheric radiative losses. \citet{Straus+etal2008} detected low-frequency gravity waves in the Sun's atmosphere, and claimed that their amount of energy is comparable to the radiative losses of the chromosphere. On the other hand, the simulated acoustic flux obtained by them is a factor of 10 smaller, consistent with \citet{Fossum+Carlsson2005}. Regarding the Alfv\'en waves, \citet{dePontieu+etal2007} claim to detect them in spicules in the upper chromosphere and estimate the energy flux carried by these waves to be sufficient to accelerate the solar wind and possibly to heat the quiet corona.

Most of the works that address the problem of upper atmosphere heating by waves refer to quiet Sun regions. In this paper we aim to study the energy balance in a sunspot umbra. An earlier work in this topic is the one by \citet{Kneer+etal1981}, who analize the time lags and rms velocities between the lines \NaIDone\ and \NaIDtwo\ and found that the waves supply to the chromosphere an energy flux of about $5\times 10^4$ erg cm$^{-2}$ s$^{-1}$, which is much lower than the chromospheric radiative losses of about 10$^6$ erg cm$^{-2}$ s$^{-1}$. Our approach consists in using 3D numerical simulations to reproduce the oscillatory pattern from the photosphere to the chromosphere obtained from temporal series of spectropolarimetric data. The resemblance between the simulated and observed waves allows us to extract conclusions from the numerical calculations provided that they correspond to a realistic phenomenon. In Section \ref{sect:procedures} we describe the observational data and the numerical method employed. Sections \ref{sect:mhs_model} and \ref{sect:driver} discuss the construction and properties of the magnetohydrostatic (MHS) model of sunspot and the procedure used to introduce the driver, respectively. The analysis of the simulations and their comparison with the observations is performed in Section \ref{sect:analysis}, while in Section \ref{sect:energy} we use these results to evaluate the energy balance. Finally, the discussion and conclusions are presented in Section \ref{sect:conclusions}.

\section{Observational and numerical procedures}
\label{sect:procedures}

The observational data consists of a temporal series of co-spatial and simultaneous data obtained with two different instruments, the POlarimetric LIttrow Spectrograph
\citep[POLIS,][]{Beck+etal2005b} and the Tenerife Infrared
Polarimeter II \citep[TIP-II,][]{Collados+etal2007}, both
attached to the German Vacuum Tower telescope at the Observatorio
del Teide at Tenerife on August 27$^{th}$ 2007. TIP-II provides Stokes spectra of the 10830 \AA\ region, including the \SiI\ $\lambda$ 10827 \AA\ and \HeI\ $\lambda$ 10830 \AA\ spectral lines, with a spatial sampling of $0''18$ per pixel. The blue channel of POLIS yields intensity spectra of the \hbox{\CaIIH\ $\lambda$ 3968 \AA\ } line and some photospheric \FeI\ line blends in its wings, with a spatial sampling of $0''29$ per pixel. The slit was placed over the center of a sunspot located near the center of the Sun. The thorough analysis of this dataset was presented in \citet{Felipe+etal2010b}. 

We have performed 3D numerical simulations to reproduce the observed wave pattern from the photosphere to the chromosphere. The numerical method is described in detail in \citet{Felipe+etal2010a}. The code solves the nonlinear MHD equations for perturbations. A MHS model of sunspot is perturbed with a driver force in the momentum equation. The properties of the MHS model and the driver used in the current work, are discussed in the following sections. A perfectly matched layer (PML) boundary condition \citep{Berenger1994} is applied to all boundaries in order to avoid wave reflections. The radiative energy losses are implemented following Newton's cooling law:
\begin{equation}
\label{eq:newton_cooling}
Q_{rad}=-c_v\frac{T_1}{\tau_R},
\end{equation}
\noindent where $T_1$ is the perturbation in the temperature, $\tau_R$ is the radiative relaxation time, and $c_v$ is the specific heat at constant volume. Following \citet{Spiegel1957}, $\tau_R$ is given by
\begin{equation}
\label{eq:spiegel}
\tau_R=\frac{\rho c_v}{16\chi\sigma_R T^3},
\end{equation}
\noindent where $\chi$ is the mean absortion coefficient and $\sigma_R$ is the Stefan-Boltzmann constant. This expression is valid at photospheric heights, but not at the chromosphere as it was derived by Spiegel in the approximation of local thermodynamic equilibrium. As the values of $\tau_R$ given by the Spiegel formula are not correct at chromospheric heights, we took the freedom to modify them in order to mimic the low value of $\tau_R=10$ s obtained from the observational analysis of the wave propagation at chromospheric heights in \citet{Felipe+etal2010b}.

\section{MHS model of the sunspot}
\label{sect:mhs_model}
The MHS model is constructed following the method by \citet{Khomenko+Collados2008}. These authors developed a method to calculate a thick sunspot structure in magnetostatic equilibrium with distributed currents, \ie, showing a continuous variation of field strength and gas pressure across the spot, from sub-photospheric to chromospheric layers. In current-distributed models, the field decreases from its value at the axis of the sunspot to almost zero at large radial distances. These models are constructed by combining the advantages of two different methods: in the deep layers, the magnetic topology is set and the thermodynamic variables are forced to match with this structure \citep{Schluter+Temesvary1958, Low1975, Low1980}, constructing the so-called ``self-similar'' models; at photospheric heights and above, the pressure distribution is prescribed as boundary condition at the axis of the sunspot and in the distant quiet Sun atmosphere and in between the pressure and magnetic field are iteratively changed until the system reaches an equilibrium state \citep{Pizzo1986}.

The stratification of thermodynamic variables in the photosphere (needed for the boundary conditions of the above methods) was retrieved from the inversion of the \SiI\ Stokes spectra in the umbra and the quiet Sun atmosphere, averaged in time and space, using SIR \citep{RuizCobo+delToroIniesta1992}. From the inversion of the \SiI\ Stokes profiles we can retrieve the stratification of magnetic field and thermodynamic variables up to a height of almost 800 km in the atmosphere of the observed sunspot and its surroundings. Since the numerical wave calculations need the distribution of gas pressure deeper below the photosphere as well as at higher chromospheric layers, we have smoothly joined our estimation of the photospheric variables with other models from the literature. As field-free atmosphere we used model S of \citet{Christensen-Dalsgaard+etal1996} at deeper layers and the VAL-C model of the solar chromosphere \citep{Vernazza+etal1981} in the upper layers. At the axis of the sunspot we use the \citet{Avrett1981} model in the upper layers, while the deep layers were extracted from a model by \citet{Kosovichev+etal2000} obtained from helioseismic inversions of the phase speed in sunspots. The resulting sunspot atmosphere was shifted down \hbox{350 km} in the vertical direction in order to account for the estimated Wilson depression (according to horizontal pressure balance). These stratifications were set as boundary conditions at the quiet Sun and at the umbral axis atmosphere, respectively.

The method also requires to characterize the horizontal variations of the magnetic field of the structure at some height. At each spatial position of our observations (along the slit) we have averaged the Stokes profiles of the \SiI\ $\lambda$ 10827 \AA\ line in time, and we have inverted the resulting Stokes vectors with SIR. From a gaussian fit to the spatial variation of the magnetic field obtained from the inversion at the formation height of the silicon line we have obtained the parameters which define the radial distribution of the magnetic field, \ie, its strength at the axis and the effective diameter. 

\begin{figure}[!ht] 
 \centering
 \includegraphics[width=9cm]{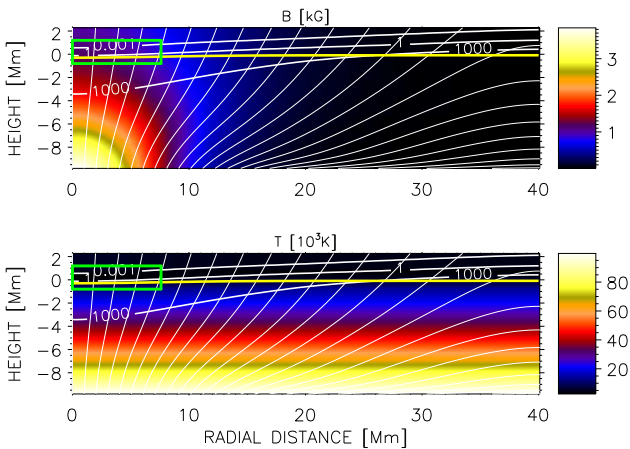}
  \caption{MHS sunspot model constructed to match the observed properties. Top: Magnetic field strength; bottom: Temperature. White thin lines are magnetic field lines. White thick lines with labels are the contours of $c_S^2/v_A^2$. The yellow line is the layer with optical depth unity. The height $z=0$ Mm corresponds to the height where optical depth is unity at quiet Sun atmosphere.}
  \label{fig:spot_BT}
\end{figure}

\begin{figure}[!ht] 
 \centering
 \includegraphics[width=8.5cm]{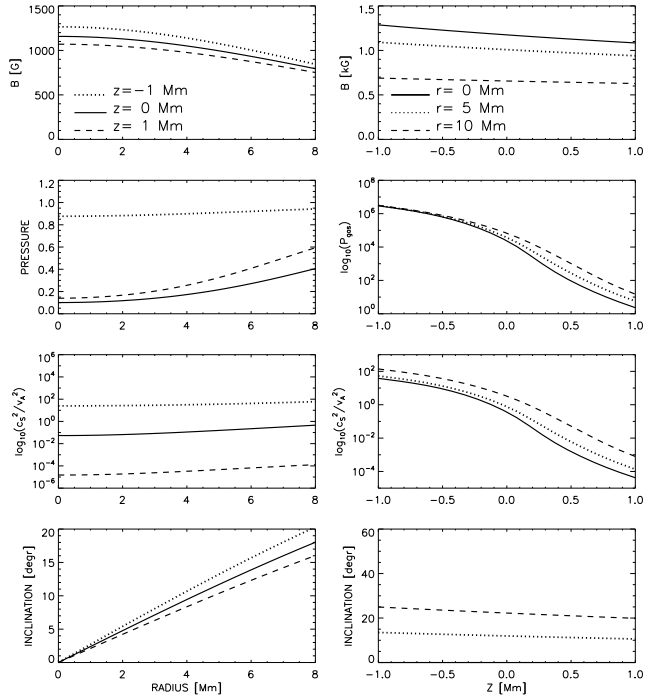}
  \caption{Distribution with radial distance (left panels) and with depth (right panels) of the magnetic field strength, pressure, ratio $c_s^2/v_A^2$, and the magnetic field inclination for the obtained sunspot model. The radial pressure distributions are normalized to their values at the right boundary (non-magnetic). Only the values corresponding to heights/distances in the green box from Figure \ref{fig:spot_BT} are shown.}
  \label{fig:spot_variables}
\end{figure}

Figure \ref{fig:spot_BT} shows the distribution of the magnetic field and temperature in the complete sunspot model. Most of the magnetic field is concentrated around the axis of the model, inside a radius of 10 Mm, and it is weaker at larger distances from the center. The orientation of the magnetic field lines changes very fast from being vertical at the axis to almost horizontal at radial distances larger than 30 Mm below $z=0$ km. Above this layer the magnetic field lines spread increasing their inclination with the distance to the axis. The properties of the model are shown in more detail in Figure \ref{fig:spot_variables}, where the distribution with radius at several heights (left panels) and the stratification with height at different distances from the axis of the sunspot (right panels) are plotted. Note that we only show values corresponding to the green box in Figure \ref{fig:spot_BT}, as this box is used in the simulations. At the axis of the sunspot the magnetic field drops from 1300 G at \hbox{$z=-1$ Mm} to \hbox{1.1 kG} at \hbox{$z=1$ Mm}. It is vertical at \hbox{$r=0$ Mm} and its inclination at photospheric layers increases with the radius, reaching an inclination around 20$^o$ at the rightmost point of the simulation domain. The gas pressure has a deficit at the magnetized regions around the center of the sunspot. Below $z=0$ km this deficit decreases with depth, as can be seen from the radial distribution of gas pressure at $z=-1$ Mm, and it almost disappears at about $-2$ Mm depth. The gas pressure changes by 6 orders of magnitude from its value at $z=-1$ Mm to $z=1$ Mm. The squared ratio between the sound speed and the Alfv\'en speed has very strong variations, from $10^2$ at $z=-1$ Mm to $10^{-5}$ at $z=1$ Mm (taking the values at the axis).

\begin{figure}[!ht] 
 \centering
 \includegraphics[width=8.5cm]{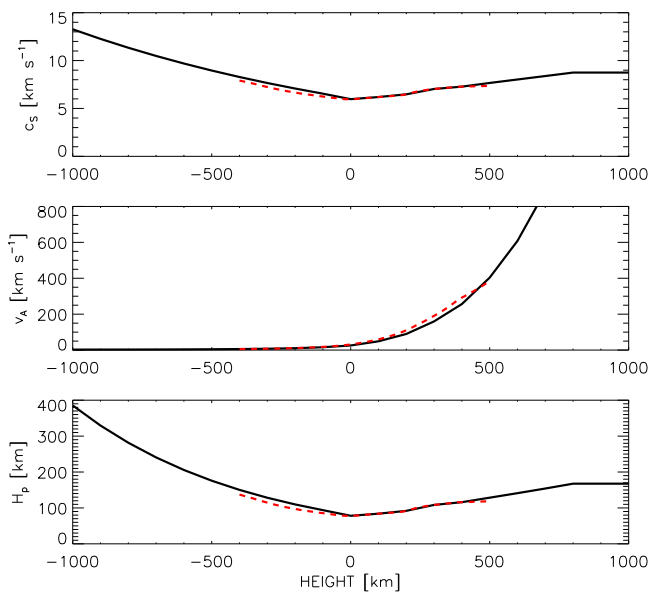}
  \caption{Comparison between the computed model of sunspot (black solid line) and the inversion (red dashed line) at the axis of the sunspot for a range of heights around the photosphere. From top to bottom: sound velocity, Alfv\'en velocity and pressure scale height.}
  \label{fig:comp_mhs_inv}
\end{figure}

It must be noted that the observed photospheric magnetic field strength
was around 2000 G, while the model has a value between 1100 G and 1300 G.
This mismatch is the result of the complex merging of all the different
model atmospheres. As explained above, the stratification of the
thermodynamic variables in the sunspot axis has been obtained by merging
three independent models: the one by Kosovichev et al. (subphotospheric
layers), the one retrieved from the inversions (photospheric layers) and
Avrett's model (upper layers). Also, the photospheric model is formed by
the model of Christensen-Dalsgaard (subphotospheric layers), the one from
the inversions (photosphere) and VAL-C (upper layers). The different
models did not coincide at the same height for all parameters, or even,
some times, discontinuities appeared in the stratification of a given
parameter, which made necessary an interpolation at some layers to merge
them smoothly. At the end, the resulting stratifications at the sunspot
axis and field-free atmospheres did not satisfy hydrostatic equilibrium
and pressures were re-calculated to impose it. This re-calculation gave
rise to a variation of the pressure stratification at both the sunspot
axis and quiet sun atmospheres, with a reduced pressure deficit (which
turned at the end into a lower  sunspot magnetic field strength). The
later merging with Low's model to generate a deep sunspot also suffers
from the same problem, but this represented a minor correction. At this
point, we could have opted to improve the sunspot model to match the
observed magnetic field strength. But, given that the retrieved
stratifications of the parameters which mainly determine the properties of
wave propagation (Alfv\'en and sound velocities and pressure scale height
after the iterative process) are very similar to those obtained from the
inversions (see Figure \ref{fig:comp_mhs_inv}), we were confident that the model was adequate
to study the propagation of waves in the atmosphere of the observed
sunspot.

\section{Introduction of the driver}
\label{sect:driver}

In this work our aim is the reproduction of the observed wave pattern by means of numerical calculations. In this respect it is crucial to choose the most appropriate way to introduce the observed velocity as a driver. We have chosen the velocity measured with the \SiI\ line as the driver of the simulation, since it is the line which is formed the deepest over the set of lines that we have observed. At the formation height of the \SiI\ line, the numerical simulation should have a vertical velocity as close as possible to the measured LOS velocity. The photospheric oscillations are dominated by waves in the 5 minute band and, thus, the power excited in the simulation in this band must resemble the observed one. However, it is even more critical to introduce correctly the power at higher frequencies. Waves with frequencies above the cutoff propagate upward and dominate the higher layers. The wave pattern at the chromosphere will depend on the power introduced by the driver at those high frequencies as well as on the initial phase of these high frequency waves.

It is interesting to note that the simulations by \citet{Carlsson+Stein1997} showed that it is possible to derive a transfer function for accurately relating the observed velocity with a piston velocity at the bottom boundary, where the latter may be located at a deeper position. This approach is not valid in our case. \citet{ Carlsson+Stein1997} performed one dimensional hydrodynamic simulations, where they only can propagate acoustic longitudinal waves in the vertical direction, and there is a unique transfer function for this wave, without ambiguity. However, in the three dimensional case, there are three distinct MHD waves (fast, slow and Alfv\'en), and the transfer function cannot be defined in an univocal way. Moreover, in our simulations, unlike those of \citet{Carlsson+Stein1997}, between the bottom boundary and the formation height of the \SiI\ line there is the layer where sound and Alfv\'en velocities are equal. There, the different modes mix up, which makes it impossible to find a unique transfer function. For all these reasons, we have decided to impose the driver at the formation height of the \SiI\ line, avoiding any hypothesis about the propagation at deeper layers.

Several strategies may be developed to that aim. On the one hand, one can set the observed oscillations as a boundary condition in a computational domain where the bottom boundary coincides with the formation height of the \SiI. On the other hand, one can calculate the force which corresponds to the measured velocity and introduce it directly into the equation of motion \citep[\eg,][]{Felipe+etal2010a}. 

In the case of the first approach, several problems arise. It is not valid just to set the vertical velocity, since it is necessary to impose at the bottom boundary the fluctuations of all the variables self-consistently. From the inversion of the Stokes profiles we can retrieve the variations of all these magnitudes, but it is difficult to obtain reliable values with a good spatial and time resolution related to a single layer in geometrical height, not optical depth \citep[see][]{RodriguezHidalgo+etal2001}. Another option is to calculate the polarization relations of all the variables which agree with the vertical velocity measured from the Doppler shift, but it is a tough work in such a realistic case. 

\begin{figure}
\centering
\includegraphics[width=8.5cm]{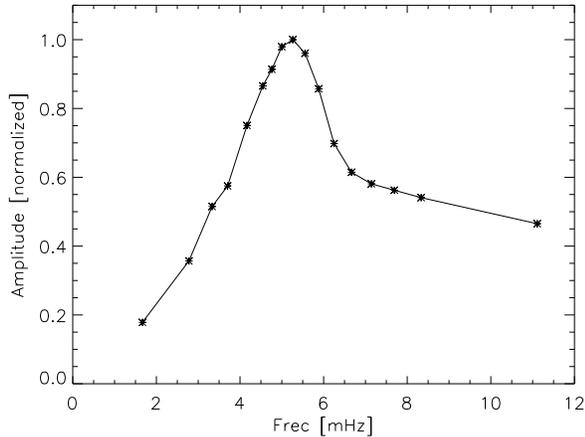}
\caption{Variation of the amplitude of the velocity obtained at the stationary stage of simulations with the frequency of the force driver, normalized to the maximum value.} \label{fig:ampl_vs_frec}
\end{figure}

We found thus more convenient to introduce the retrieved force as a source function $S_z(x,z_{Si},t)$ in the momentum equation. This driver introduces mechanical energy in the system, and consequently, the energy equation would also have to be modified. However, in this particular simulation, we use the internal energy instead of imposing conservation of the total energy. We have omitted the introduction of an additional term in the internal energy equation since the energy of the driver should not be employed in heating the plasma. The vertical force has the following dependence on the velocity:

\begin{equation}
\label{eq:Semp}
S_{z}(x,z_{Si},t)=\int_{\nu_{0}}^{\nu_{1}} v_{z,obs}(x,\nu,z_{Si})\frac{1}{A(\nu)}\Delta(\nu)e^{2\pi i \nu t}d\nu,
\end{equation}

\noindent where $v_{z,obs}(x,\nu ,z_{Si})$ is the observed with the \SiI\ $\lambda$ 10827 \AA\ line velocity amplitude at frequency $\nu$ for all the spatial positions $x$ along the slit, and the functions $A(\nu)$ and $\Delta(\nu)$ include the dependence of the force amplitude and phase delay on frequency. The form of $A(\nu)$ was evaluated by analyzing numerically the response of the atmosphere to a force driver with a fixed frequency. We have carried out a set of simulations using a harmonic driver of fixed $\nu$ located at the photosphere with the force amplitude independent of $\nu$. The amplitude of the velocity retrieved after reaching the stationary regime of the simulations is chosen as the response of the atmosphere to a harmonic wave. Figure \ref{fig:ampl_vs_frec} shows the values obtained for all the simulations performed, normalized to the maximum amplitude. The amplitude $A(\nu)$ increases almost linearly from $\nu=1.9$ mHz to around 5.2 mHz, and for higher frequencies it decreases as the inverse of the frequency. Its maximum is located at a frequency close to the local cutoff frequency. The particular form of this dependence varies with the parameters of the sunspot atmosphere, and it critically depends on the location of the driver in height. We found that a change in the height location of the driver shifts the frequency of the maximum response. At those frequencies for which the response of the velocity to the force is very efficient (for example, at 5.3 mHz) the factor which multiplies the velocity to obtain the force from Equation (\ref{eq:Semp}) must be lower than for those frequencies with a poorer response (for example, at 3 mHz). For this reason $A(\omega)$ appears in the denominator in Equation (\ref{eq:Semp}).

To get an idea about the form of $\Delta(\nu)$ (phase response), we have carried out a simulation using Equation (\ref{eq:Semp}) to introduce the observed driver in the simulations, including the expression of $A(\nu)$ from Figure \ref{fig:ampl_vs_frec} but ignoring the phase dependence $\Delta(\nu)$. The phase difference between the observed velocity and the one used in the simulation at the height where the driver was introduced shows a phase delay of $\Delta \phi=\pi/2$ for frequencies below 4 mHz and $\Delta \phi=0$ for frequencies higher than 6 mHz, with an almost linear variation between 4 and 6 mHz. The function $\Delta(\omega)$ has been constructed in order to characterize this behavior. As the velocity at low frequencies needs a quarter of a period to account for the variations of the force, at these frequencies $\Delta(\nu)$ shifts the source driver backwards. We have set $\nu_{0}=1.5$ mHz and $\nu_{1}=20$ mHz.

\section{Analysis of the simulations}
\label{sect:analysis}

\subsection{Set up}
\label{sect:setup}
We have introduced the driver from Equation (\ref{eq:Semp}) as a force perturbation in the MHS model of the observed sunspot obtained in Section \ref{sect:mhs_model}. The height where the optical depth at 5000 \AA\ is unity at the quiet Sun atmosphere was chosen as $z=0$ Mm. According to \citet{Bard+Carlsson2008}, in a model of sunspot the \SiI\ line forms at a geometrical height of 308 km above the height where continuum $\tau_{5000}=1$, which corresponds to $z=-42$ km for the adopted Wilson depression of \hbox{350 km} where the zero-height level is defined by the non-magnetic atmosphere. We have to take into account that we need at least \hbox{$0.9-1$ Mm} of atmosphere above the location of the driver to reproduce the travel of the wave from the \SiI\ line to the \HeI\ line, according to the geometrical height difference between these two layers \citep{Centeno+etal2006, Felipe+etal2010b}. At those high layers, the Alfv\'en speed of the sunspot MHS model takes very high values, which requires an extremely small time step in the simulations. In order to save computational time and avoid problems with the top PML boundary, we have located the driver slightly deeper, at $z=-100$ km. At this height the ratio $c_S^2/v_A^2$ is around 0.8. Since we are introducing a vertical force in a magnetically dominating region with an almost vertical magnetic field, it mainly drives longitudinal waves which correspond to a slow magneto-acoustic mode.

In order to compare the numerical simulation with the observational data, we have assigned a fixed $z$ to the formation height of each spectral line, and we have assumed that the vertical velocity at that location corresponds to the velocity measured from the Doppler shift of the line. Following the formation heights retrieved from \citet{Felipe+etal2010b} and taking into account that the velocity obtained from the observations of the \SiI\ line was imposed as a driver at $z_{Si}=-100$ km, the layers of the computational domain selected as representative of the heights of formation of the rest of the spectral lines are $z_{Fe}=175$ km, $z_{Ca}=600$ km and $z_{He}=725$ km for the \FeI\ $\lambda$ 3969.3 line, \CaIIH\ core, and \HeI\ line, respectively.

In the horizontal directions the computational domain cover $14.8\times 8.4$ Mm, with a spatial steps of $\Delta x=\Delta y=100$ km. In the vertical direction it spans from $z=-0.6$ Mm to $z=1$ Mm, excluding the PML layer, with a spatial step of $\Delta z=25$ km. We set the force $S_z(\nu,x,z_{Si},t)$ for all the heights inside a layer of a chosen thickness in the $y$ and $z$ directions, but smoothly modulated to zero after a few grid points, and covering the slit of the observations in the $x$ direction. In this manner, the driver only acts in a narrow layer around the formation height of the silicon line for an elongated region in the $x$ direction.

\subsection{Oscillations at the formation heights of the spectral lines}

\begin{figure}
\centering
\includegraphics[width=9cm]{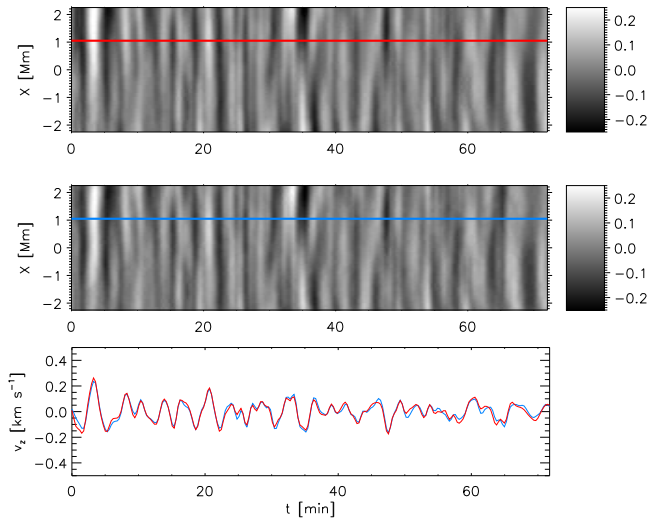}
\caption{Velocity maps of the \SiI\ line (the vertical direction is the direction along the slit, the horizontal
direction is time). Top: Observational, measured from the Doppler shift of the \SiI\ line; middle: numerical, vertical velocity at the formation height of the \SiI\ line; bottom: comparison of the observational (red line) and numerical (blue line) velocity at x=1.1 Mm.} \label{fig:si_maps}
\end{figure}

Figure \ref{fig:si_maps} shows a comparison between the LOS velocity map of the umbra observed with the \SiI\ line (top panel) and the vertical velocity of the simulation at the height where the driver was introduced (middle panel). Negative velocities (appearing as black shaded regions) indicate upflows, where the matter moves toward the observer, while white regions are downflows. Both wave patterns are almost identical, with an amplitude below \hbox{0.3 km s$^{-1}$}, indicating that the evaluation of the parameter $A(\nu)$ and $\Delta(\nu)$ for Equation (\ref{eq:Semp}) is correct. The bottom panel shows the time evolution of the velocity at a certain location inside the umbra, confirming once again that the simulation fits well the observations.

\begin{figure}
\centering
\includegraphics[width=9cm]{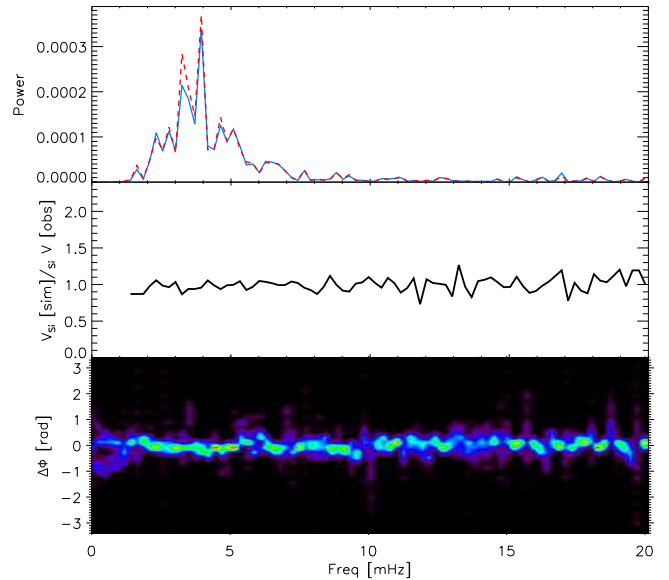}
\caption{Top: Power spectra of the observed \SiI\ velocity (red dashed line) and the simulated velocity at the height where the driver is introduced (blue solid line); middle: ratio of the simulated amplitude to the observed one; bottom: differences in phase.} \label{fig:si_spectra}
\end{figure}

To perform a more detailed comparison between the velocities at the height of the driver, we show in Figure \ref{fig:si_spectra} a spectral analysis of the simulation and the observation. The top panel illustrates the power spectra of the observed velocity (red dashed line) and the numerical vertical velocity at the height of the driver (blue solid line) averaged over slit. The ratio between the amplitudes of both velocities (simulated/observed) is given in the middle panel. In the whole frequency range the ratio is around unity, indicating a good match between the driver and the real oscillation. The power peak at 3 mHz is a bit lower in the simulation. For frequencies above 7 mHz, where the power is very low, the ratio departs slightly from unity, varying between 0.7 and 1.3. The bottom panel of Figure \ref{fig:si_spectra} shows the phase difference between the measured and simulated velocities. At each frequency we have calculated histograms of the phase difference in all the spatial points inside the umbra. The color scale of the panel indicates the relative occurrence of a given phase shift, spanning from black (low) to red (high). Negative phase difference means that the simulated velocity lags the observed one. Both oscillatory signals are in phase for all frequencies.

Now we can compare the observed velocities obtained from different spectral lines with the simulated ones at heights defined in Section \ref{sect:setup}.

\begin{figure}
\centering
\includegraphics[width=9cm]{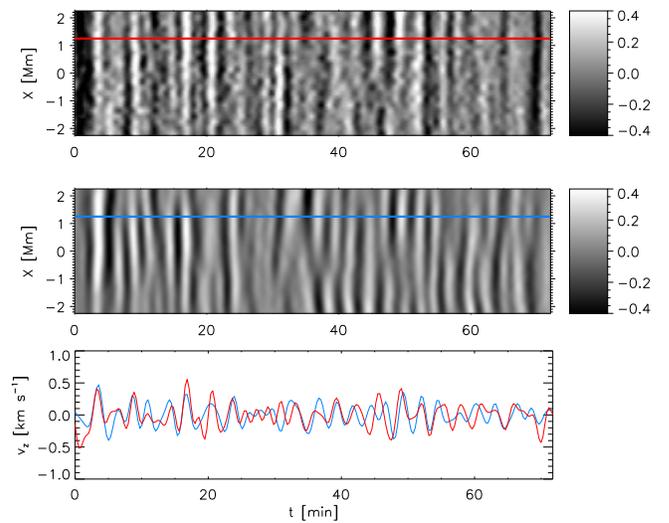}
\caption{Velocity maps of the \FeI\ $\lambda$ 3969.3 line. Top: Observational, measured from the Doppler shift of the \FeI\ line; middle: numerical, vertical velocity at the formation height of the \FeI\ line; bottom: comparison of the observational (red line) and numerical (blue line) velocity at x=1.1 Mm.} \label{fig:fe_maps}
\end{figure}

The \FeI\ line is formed about 280 km above the \SiI\ line, where the driver was introduced. Figure \ref{fig:fe_maps} shows the velocity map observed with the \FeI\ $\lambda$ 3969.3 line and the simulated one at the corresponding height. The Doppler velocity map retrieved from the \FeI\ line is quite noisy, especially in this region of the umbra where the spectral line intensity is very low \citep{Felipe+etal2010b}. However, the strongest wavefronts, with an amplitude of almost 0.5 km s$^{-1}$, can be recognized and compared with the simulated wave pattern, showing a good agreement. For example, the wavefronts around $t=20$ min or the ones between $t=47$ and $t=55$ min can be clearly identified in the simulation, with a similar amplitude (bottom panel of Figure \ref{fig:fe_maps}). 

\begin{figure}
\centering
\includegraphics[width=8.5cm]{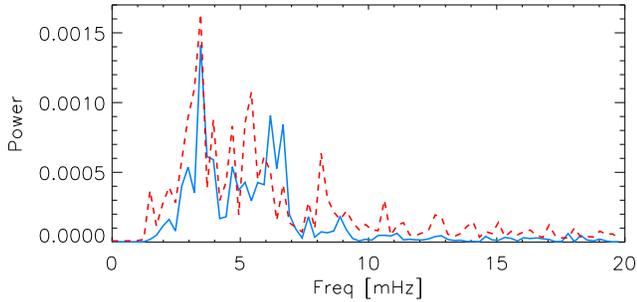}
\caption{Power spectra of the observed \FeI\ $\lambda$ 3969.3 velocity (red dashed line) and the simulated velocity at its corresponding height (blue solid line), averaged over the umbra.} \label{fig:fe_spectra}
\end{figure}

The power spectrum of the simulation at the formation height of the \FeI\ $\lambda$ 3969.3 line is given in Figure \ref{fig:fe_spectra}. The numerical simulation reproduces very well the power peak at 3 mHz, while the frequency of the secondary power peak is shifted from 5.4 mHz in the observations to 6.2 mHz in the simulations. At frequencies above 8 mHz the observational power spectra shows higher power, but we suspect that this high frequency power corresponds to the noise present in the velocity signal of the \FeI\ lines. 

\begin{figure}
\centering
\includegraphics[width=9cm]{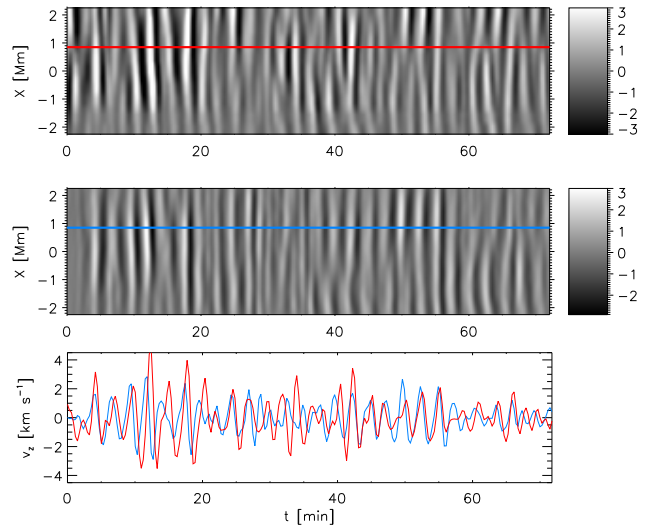}
\caption{Velocity maps of the \CaIIH\ core. Top: Observational, measured from the Doppler shift of the \CaIIH\ line; middle: numerical, vertical velocity at the formation height of the \CaIIH\ line; bottom: comparison of the observational (red line) and numerical (blue line) velocity at x=0.9 Mm.} \label{fig:ca_maps}
\end{figure}

Figure \ref{fig:ca_maps} illustrates the observational and numerical velocity maps in the case of the LOS velocity measured with the \CaIIH\ core. The \CaIIH\ core is formed at the chromosphere, and waves have propagated upwards about 700 km from the formation height of the \SiI\ line in order to reach this layer. Note that in the simulated velocity map (middle panel) the velocity signal is almost null during the first 2 minutes, due to the time spent by the slow magneto-acoustic waves to cover the distance between the driver and this height travelling at the sound speed. During this travel the period of the waves is reduced to around 3 minutes and their amplitude increases, reaching peak-to-peak values of almost \hbox{8 km s$^{-1}$}. The bottom panel of \hbox{Figure \ref{fig:ca_maps}} shows that the oscillations develop into shocks. This behavior is well reproduced by the numerical simulation. The simulated velocity map reproduces reasonably well the observed oscillatory pattern. Only in the temporal lapse between $t=27$ and $t=40$ min the simulated pattern differs significantly from the observations. Most of the observed wavefronts can be identified in the simulation, although their spatial coverage of the umbra can be slightly different. For example, in the observations at $t=50$ min a wavefront shaded in white covers from the limit of the plotted velocity map at $x=2$ Mm to around $x=0$ Mm, while in the simulations it extends from $x=2$ Mm to almost $x=1$ Mm. These differences may be due to the limitations in the configuration of the numerical simulation: on the one hand, the MHS atmosphere is an axisymmetric model, which is obviously not the case of the real sunspot. Thus, the distance travelled by the waves along the field lines may be different, producing a phase lag. On the other hand, we have introduced the driver only in a region of the umbra along the slit of the observations, and we have ignored the driving of waves in the rest of the (non observed) umbra. 

During the first 20 minute of the simulations there is some phase shift with respect to the observations, and the amplitude of the simulations is lower. Note that this time lag is evident in the wavefronts with the largest amplitude, where the nonlinearities are clear. We will discuss this behavior in Section \ref{sect:conclusions}.

\begin{figure}
\centering
\includegraphics[width=8.5cm]{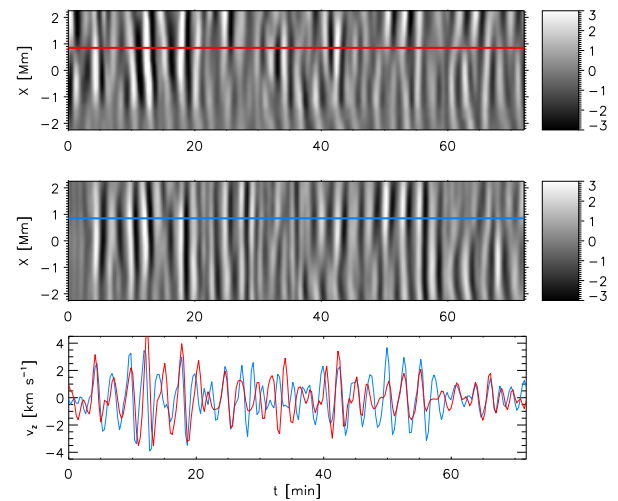}
\caption{Velocity maps of the \HeI\ line. Top: Observational, measured from the Doppler shift of the \HeI\ line; middle: numerical, vertical velocity at the formation height of the \HeI\ line; bottom: comparison of the observational (red line) and numerical (blue line) velocity at x=0.9 Mm.} \label{fig:he_maps}
\end{figure}

The spectral line formed the highest observed is the \HeI\ line, which is formed around 100 km above the \CaIIH\ core. The comparison between the Doppler velocity of this line and the vertical velocity of the simulation at the corresponding height is given in Figure \ref{fig:he_maps}. Similar to the case of the \CaIIH, most of the wavefronts of the observations can be clearly identified in the simulations, except in the temporal range between $t=27$ and $t=40$ min. The match between the observed and simulated amplitudes is also remarkable. Both maps seem to be almost in phase. There is some phase delay which coincides with the strongest shocks, but it is smaller than the one obtained for the \CaIIH\ core.

\begin{figure}
\centering
\includegraphics[width=9cm]{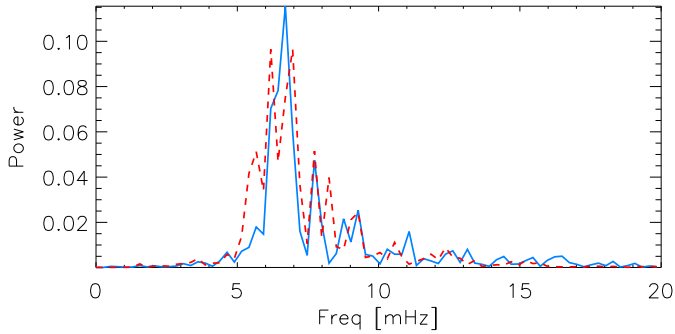}
\caption{Power spectra of the observed \HeI\ velocity (red dashed line) and the simulated velocity at its corresponding height (blue solid line).} \label{fig:he_spectra}
\end{figure}

Only those waves with frequency above the cutoff can reach the chromosphere. The increase of the amplitude of these waves with height is larger than that of the evanescent low frequency waves, and the power spectra at the chromosphere is dominated by the peak at 6 mHz. For example, in the case of the power spectra of the \HeI\ line (Figure \ref{fig:he_spectra}), both the observations and simulation have their power concentrated around this frequency. The observational power has three power peaks in the 3 minute band, located at 5.5, 6 and 7 mHz. The simulated power is concentrated at a single peak between the two highest peaks of the observations. The simulated peak at 5.5 mHz is lower than the observed one. The simulations also reproduce the power peaks at 7.7 mHz and 9 mHz, and the low power at frequencies below 5 mHz. At frequencies above 13 mHz the simulated power is larger than the observational one.

\subsection{Propagation from the photosphere to the chromosphere}

The propagation properties of waves are analysed by means of the phase difference ($\Delta \phi$) and amplification spectra. The value of $\Delta \phi$ gives the time delay between the
oscillatory velocity signals from two spectral lines. We assume
that the difference between them is mainly due to the difference
of the formation height of the two lines. The amplification spectra are calculated as the ratio between the wave power at two layers. They give us information about variations of oscillation amplitude with height. The estimation of the statistical validity of the amplification and phase difference spectra can be done by calculation of the coherence spectra. This calculations for the observed dataset were presented previously in \citet{Felipe+etal2010b}.

In order to reproduce correctly the phase and amplification spectra, the introduction of the term $Q_{rad}$ in the energy equation for the simulations (see Section \ref{sect:procedures}), to take into account the energy losses, has proved to be fundamental. This is so due to several reasons. Firstly, the radiative losses produce some damping of the waves, reducing their amplitude. Secondly, the cutoff frequency of the waves is also affected by radiative transfer \citep{Roberts1983,Khomenko+etal2008c}. When the radiative timescale $\tau_R$ is small enough, the cutoff frequency is expected to decrease, compared to the adiabatic case. Finally, the inclusion of radiative losses also produces an increase of the phase difference compared to adiabatic case \citep[see Figure 7 from][]{Centeno+etal2006}.

\begin{figure}
\centering
\includegraphics[width=9cm]{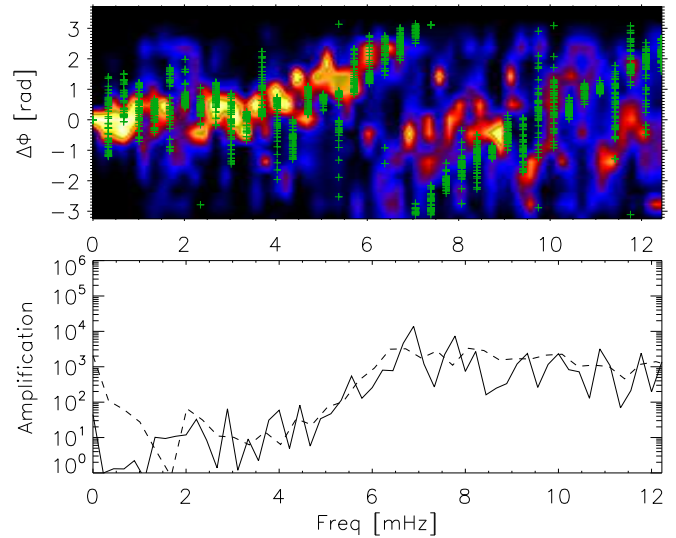}
\caption{Top: Phase difference spectra between \SiI\ and \HeI\ velocities in the umbra.  The color code shows the relative occurence of a given phase shift in the observations. The green crosses are the results of the simulation for all the spatial points. Bottom: Amplification spectra for the observation (black solid line) and the simulations (black dashed line).} \label{fig:spiegel_sihe}
\end{figure}

Figure \ref{fig:spiegel_sihe} shows the phase difference and amplification spectra between the photospheric \SiI\ line and the chromospheric \HeI\ line, for both observational and simulated velocities. From 1 to 7 mHz, where the coherence of the observations is high, the simulated phase difference precisely matches the observed one, with a null phase difference for frequencies below 4 mHz and an almost linear increase between 4 and 7 mHz. At higher frequencies the coherence of the observed phase difference is lower, but the simulated one keeps its linear increase. With regards to the amplification spectra, for frequencies above 1.5 mHz the simulated spectra reproduces properly the observed one. The smallest frequencies show a high numerical amplification, which is possibly due to the difficulties of the PML to damp these long period waves. The thickness of the PML layer should be proportional to the wavelength of the wave that must be absorbed, and the employed PML is obviously not optimized for such long period waves. However, since the power at these low frequencies is small, their overall influence on the simulations is negligible.

\begin{figure}
\centering
\includegraphics[width=9cm]{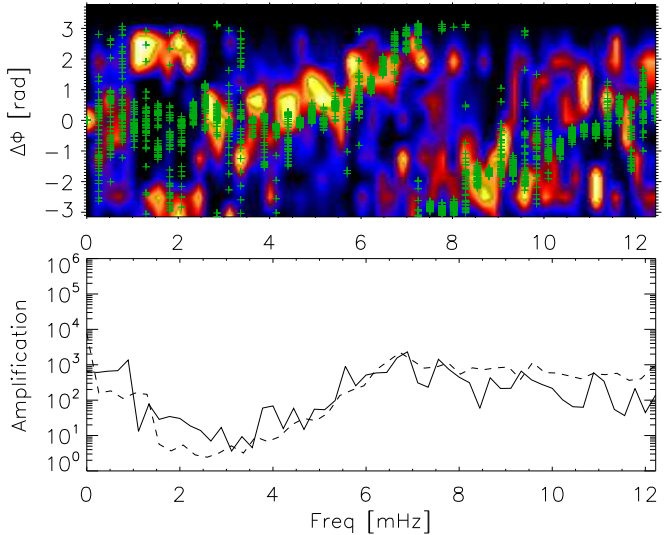}
\caption{Top: Phase difference spectra between \SiI\ and \CaIIH\ velocities in the umbra.  Bottom: Amplification spectra. The format is the same as in Figure \ref{fig:spiegel_sihe}.} \label{fig:spiegel_sica}
\end{figure}

A similar result is found between the velocity obtained with the \SiI\ and the \CaIIH\ core (Figure \ref{fig:spiegel_sica}), since the latter is formed just around 100 km below the \HeI\ line. The simulated phase difference is zero for frequencies below 4 mHz, and it increases at higher frequencies. It matches the observed phase shift between 2.5 and 9 mHz. At higher frequencies the observed phase difference is noisier, while in the frequency range between 1 and 2.5 mHz it takes a value of $\pi$ or $-\pi$. In the analysis of the observations we expressed our doubts about the reliability of this phase shift, so we are not surprised that the simulation does not reproduce it \citep[see][]{Felipe+etal2010b}.

\begin{figure}
\centering
\includegraphics[width=9cm]{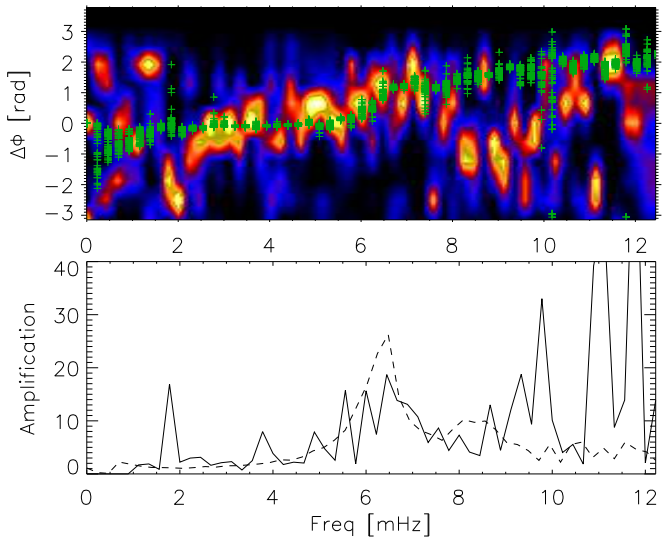}
\caption{Top: Phase difference spectra between \SiI\ and \FeI\ $\lambda$ 3969.3 velocities in the umbra.  Bottom: Amplification spectra. The format is the same as in Figure \ref{fig:spiegel_sihe}.} \label{fig:spiegel_sife}
\end{figure}

Figure \ref{fig:spiegel_sife} shows the phase and amplification spectra between two photospheric lines, the \SiI\ and the \FeI\ $\lambda$ 3969.3 lines. The observed phase difference has high coherence between 2 and 8 mHz \citep[Figure 12 in][]{Felipe+etal2010b}, and in this frequency range the simulated phase delay fits the observational one reasonably well, showing a $\Delta \phi =0$ for frequencies below 4 mHz and an increase for the higher frequencies. For frequencies below 2 mHz and above 8 mHz the observed phase difference spreads out and has lower coherence. The behavior of the simulated amplification spectra is similar to the observed one between 0 and 8 mHz, but the numerical amplification is larger at the peak around 6.5 mHz. Note also that around 3.5 mHz the observed amplification shows a peak, not reproduced in simulations. This was previously seen in the power spectra in Figure \ref{fig:fe_spectra}. At higher frequencies, the observed amplification has some high peaks, but they are not trustable due to the poor quality of the \FeI\ velocity map.

\begin{figure}
\centering
\includegraphics[width=9cm]{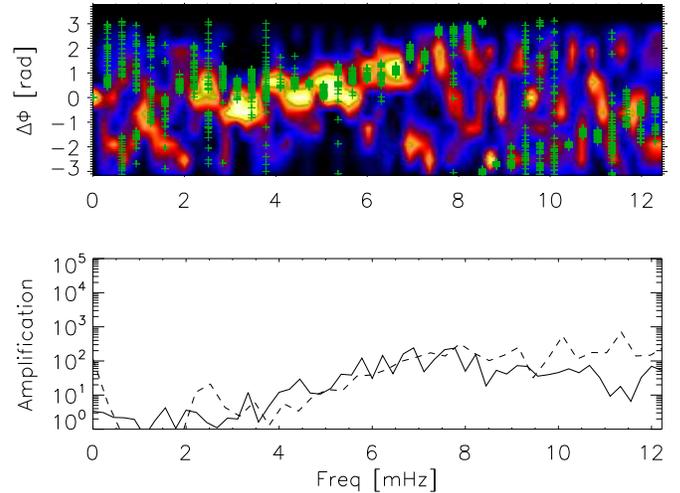}
\caption{Top: Phase difference spectra between \FeI\ $\lambda$ 3969.3 and \HeI\ velocities in the umbra.  Bottom: Amplification spectra. The format is the same as Figure \ref{fig:spiegel_sihe}.} \label{fig:spiegel_fehe}
\end{figure}

The spectra between \FeI\ $\lambda$ 3969.3 and \HeI\ lines are shown in Figure \ref{fig:spiegel_fehe} and are similar to those in Figures \ref{fig:spiegel_sihe} and \ref{fig:spiegel_sica}, since all of them correspond to the pairs of lines including one photospheric and one chromospheric line. The numerical phase and amplification spectra match the observational ones in the frequency range between 2 and 8.5 mHz. Out of this range the observational phase spectra are very noisy and with a lower coherence, meaning that these phase shifts are not reliable. At frequencies above 8 mHz the observational amplification seems to decrease, but this tendency is not reproduced by the simulated one.

\section{Energy balance}
\label{sect:energy}

The driver is located just above the layer where the Alfv\'en speed is equal to the sound speed. At this height the $c_s^2/v_A^2$ parameter is 0.8. Since our driver is a vertical force, it mainly generates oscillations in the vertical velocity. Inside the umbral photosphere the magnetic field is almost vertical, and most of the energy introduced by the vertical force has an acoustic nature, which corresponds to a slow mode in this low-$\beta$ region. Because of the vertical thickness of the driver, some part also acts at $c_S=v_A$ or below, and it produces some fast magneto-acoustic waves in the region just below the layer $c_S=v_A$. These waves are partially transformed into a fast magnetic mode above the height where $c_S=v_A$, which is reflected towards deeper layers due to the gradients of the Alfv\'en speed. The energy of this mode is very low. However, it must be taken into account that in this simulation we are introducing the observed velocity at the formation height of the \SiI\ line and, thus, the wave pattern below this layer is not reliable. In the real sunspot, waves propagate from deeper layers and in their upward propagation they reach the layer where $c_S=v_A$. Some significative part of the energy of these waves is transformed into fast magnetic waves at this height, and the contribution of the fast modes in the low-$\beta$ region must be larger than the one estimated in this simulation.

The slow acoustic mode generated directly by the driver propagates upwards along the field lines. According to Figure \ref{fig:si_spectra}, at $z=-100$ km most of its power is concentrated in the 5 minute band, in a frequency band between 3 and 4 mHz. At this layer the cutoff frequency is $\nu_c=4.7$ mHz, and it increases with height until it reaches the maximum value $\nu_c=6$ mHz at $z=200$ km. It means that the oscillations in the 5 minute band introduced by the driver cannot propagate upwards, and they form evanescent waves which do not supply energy to the higher layers. Only those waves with frequencies above the cutoff can propagate to the chromosphere. 

\begin{figure}
\centering
\includegraphics[width=9cm]{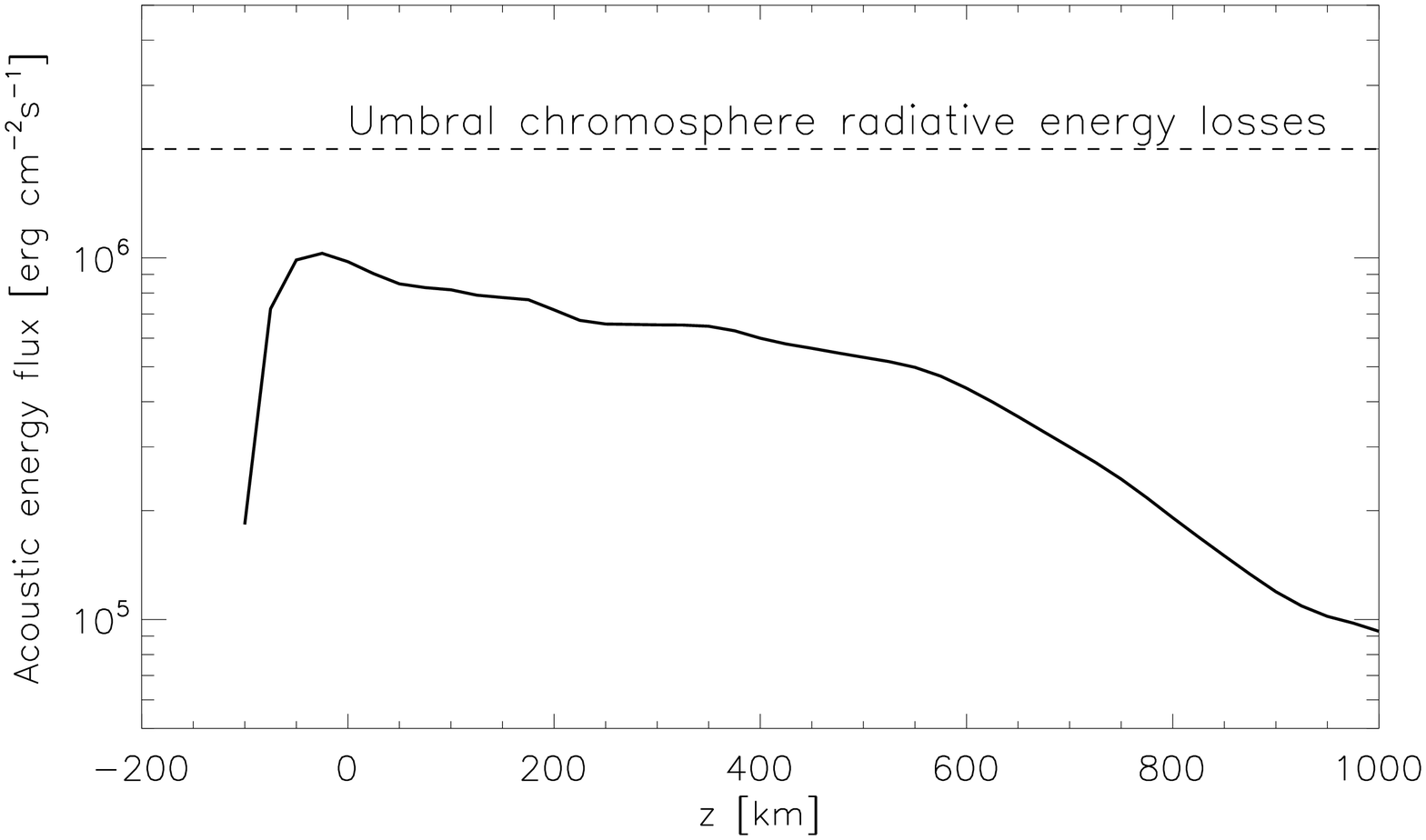}
\caption{Average acoustic energy flux inside the umbra in the simulation.} \label{fig:Fac_height}
\end{figure}

During the travel of high frequency waves to upper layers their amplitude increases due to the drop of the density. At $z=450$ km the waves develop into shocks and they reach peak-to-peak amplitudes around 8 km s$^{-1}$ at the formation height of the \CaIIH\ core ($z=600$ km) and the \HeI\ line ($z=725$ km). These waves supply energy to the chromosphere, but, is this acoustic energy enough to balance the radiative losses of the chromosphere? According to \citet{Avrett1981}, the radiative energy losses in the umbral chromosphere amount to \hbox{$2.6\times10^6$ erg cm$^{-2}$ s$^{-1}$}, while \citet{Kneer+etal1981} estimate a similar value of \hbox{$1-2\times 10^6$ erg cm$^{-2}$ s$^{-1}$}. These losses should be balanced by some energy source. Figure \ref{fig:Fac_height} illustrates the variation of the acoustic flux with height, which was calculated following \citet{Bray+Loughhead1974}:
\begin{equation}
{\bf F}_{\rm ac}=\langle p_1{\bf v}_1\rangle,
\label{eq:Fac}
\end{equation} 

\noindent where $p_1$ and ${\bf v}_1$ are the fluctuations in gas pressure and velocity, respectively. It was averaged in time for all points inside the umbra at the stationary stage of simulations. At the position of the driver (z=-100 km) it is zero because it represents the average between the upward and downward fluxes. It grows up to \hbox{10$^{6}$ erg cm$^{-2}$ s$^{-1}$} just above the driving layer and decreases with height due to the dissipation and radiative energy losses. The decrease between $z=200$ km and $z=600$ km is mainly due to the impossibility of the waves with frequencies below the cutoff frequency to propagate to higher layers. Above $z=600$ km the decrease of the acoustic flux is even more pronounced because of the dissipation produced by the shocks and the losses caused by the low radiative relaxation time at chromospheric heights (we set $\tau_R=10$ s above 600 km).

The magnetic flux associated to the fast magneto-acoustic waves can be calculated following \citet{Bray+Loughhead1974} as

\begin{equation}
{\bf F}_{\rm mag}=\langle {\bf B}_1\times({\bf v}_1\times {\bf B}_0)/\mu_0\rangle.
\label{eq:Fmag}
\end{equation}

It shows a maximum of 10$^4$ erg cm$^{-2}$ s$^{-1}$ at $z=0$ km, and it drops at upper layers with a value of $2\times10^3$ erg cm$^{-2}$ s$^{-1}$ at $z=750$ km since these waves are reflected down towards the photosphere due to the gradients of the Alfv\'en speed \citep{Rosenthal+etal2002,Khomenko+Collados2006, Felipe+etal2010a}. However, it must be taken into account that most of the energy of the vertical force that we introduce as a driver mainly goes to the slow magneto-acoustic mode, and this calculation of the flux carried by the fast mode may be underestimated.

\begin{figure}
\centering
\includegraphics[width=9cm]{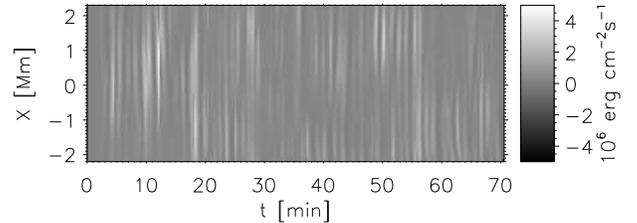}
\caption{Acoustic energy flux map at $z_{He}$. Positive values (white) indicate upward flux.} \label{fig:Fac_he}
\end{figure}

The acoustic flux shows important variations with time. For example, Figure \ref{fig:Fac_he} shows the temporal evolution of the acoustic flux inside the umbra at the formation height of the \HeI\ line. Positive values (shaded in white) represent upward flux. The wavefronts of the slow acoustic waves in the 6 mHz band are accompanied by acoustic energy which reaches almost \hbox{$5\times 10^{6}$ erg cm$^{-2}$ s$^{-1}$} for the strongest shocks. The average value of the acoustic flux at this height is \hbox{$3\times 10^{5}$ erg cm$^{-2}$ s$^{-1}$}, which is an order of magnitude below the required value to balance the radiative losses. Only at the moments when shocks reach the chromosphere the energy supplied by the slow acoustic waves is similar to the chromospheric radiative losses.

\section{Discussion and conclusions}
\label{sect:conclusions}

We have presented numerical simulations which closely reproduce the real wave propagation observed in the umbra of a sunspot from the photosphere to the chromosphere. The analysis of these observations was previously reported in \citet{Felipe+etal2010b}. As proved in \citet{Felipe+etal2010a}, our code is able to manage the propagation of waves with long realistic periods, even in the hard conditions imposed by the high Alfv\'en speeds at the sunspot chromosphere.

The similarity between the simulations and the observations is achieved in two steps. Firstly, we had to construct a MHS model of a sunspot with properties as close as possible to the observed one. We have followed the method of \citet{Khomenko+Collados2008}, introducing some modifications in order to take into account thermodynamic and magnetic properties of the particular observed sunspot, as retrieved from the inversion of the \SiI\ line. Secondly, this static model was perturbed by a force which drives a vertical velocity similar to the LOS velocity measured from the Doppler shift of the \SiI\ line. 

Once the observed velocity is introduced in the MHS atmosphere, it mainly drives slow magneto-acoustic waves in the low-$\beta$ region. Most of their power is concentrated in the 5 minute band, with frequencies between 3 and 4 mHz, and they form standing waves due to the higher cutoff frequency of the atmosphere. The driver also generates high frequency waves, which propagate upwards to the chromosphere. In their travel through the sunspot atmosphere, they reach the formation height of several observed spectral lines. The simulated velocity maps and power spectra at their corresponding formation heights reproduce reasonably well those obtained for the chromospheric \CaIIH\ core and \HeI\ line. In the case of the \hbox{\FeI\ $\lambda$ 3969.3} line, the comparison is hindered due to the poor quality of the observed velocity map. 

The comparison between the simulated and observed velocity maps at chromospheric heights reveals some phase delay at the strongest wavefronts, in the sense that the observed wavefronts lag the simulated ones. In the case of the \CaIIH\ core the phase lag is 0.4$\pi$. The delay of the observed relative to the simulated \HeI\ velocities is somewhat smaller. This delay, together with the larger amplitude of the observed strongest shocks, may indicate that in the dynamical sunspot chromosphere, the nonlinear waves shift the formation height of spectral lines to higher layers. This is especially evident for the \CaIIH\ core. In the analysis of these simulations, we have chosen a certain height representative of the layer from which the information about the velocity measured with the observed line comes. Obviously, this procedure is a simplification, since the contribution function from every spectral line spans over a thick layer. The height that we are considering corresponds to the average height derived in \citet{Felipe+etal2010b}. This approximation seems to be good for the photospheric lines, but not for the highly nonlinear chromospheric region. 

Our simulations reproduce the phase and amplification spectra between several pairs of lines with a remarkable match. For those spectra between a photospheric and a chromospheric signal, the phase difference shows stationary waves with $\Delta \phi =0$ below 4 mHz. At higher frequencies the waves progressively propagate with $\Delta \phi$ increasing with the frequency. The empirical cutoff of 4 mHz is assigned from the estimation of the frequency at which the phase difference starts to depart from $\Delta \phi =0$. In the solar atmosphere, the cutoff is a local parameter which depends on the temperature and is stratificated with height. In our MHS model its maximum value is 6 mHz, and only waves with frequencies above this value can propagate all over the atmosphere. However, in most of the atmosphere the local cutoff is below 6 mHz, and for this reason the phase difference spectra show propagation for frequencies in the range between 4 and 6 mHz.

The only strong discrepancy between observations and simulations is found in the amplification spectra at frequencies above 8 mHz (Figures \ref{fig:spiegel_sica} and \ref{fig:spiegel_fehe}). The amplification is cleary higher in the simulations than in the observations, and the observed power spectra of the chromospheric lines show lower power than the simulated equivalent (Figure \ref{fig:he_spectra}). There are several causes that could be responsible for this discrepancy. The comparison of the power spectra of the velocity retrieved from the \SiI\ line and the one obtained in the simulation at its formation height (middle panel of Figure \ref{fig:si_spectra}) shows some slight differences which are especially significant above 8 mHz. Since the amplitude of the waves increases exponentially with height, these small discrepancies could be the origin of the larger ones that appear at larger heights. For example, Figure \ref{fig:si_spectra} shows that the simulation has an excess of power at the height of the driver at 13 mHz and 17 mHz, and these frequencies also present higher power than the observations at the chromosphere (Figure \ref{fig:he_spectra}). However, some frequencies where the power of the photospheric velocity in the simulation is not larger than the power of the \SiI\ line also present a significative difference at higher layers. It is possible that most of the power generated by the driver at high frequencies (which match the power obtained in the observations) does not correspond to real photospheric oscillations, but it is introduced by some observational limitations. In this case, the simulation would propagate upward some oscillatory power which is not present in the Sun, and although it vanishes at deep layers, its contribution in the chromosphere is important. Another possible cause of the different power of the high frequency waves at the chromosphere is the simple energy exchange implemented in this simulation. A more realistic treatment could produce a stronger radiative damping of the high frequency waves.

The analysis of the energy flux carried by magnetoacoustic waves to high layers reveals that it is not enough to heat the chromosphere. The average energy supplied by slow acoustic waves is around 3 to 9 times lower than the amount required to balance the radiative losses, depending on the reference taken for the umbral chromospheric radiative losses. However, the simulation overestimates the power of the waves with frequencies between 8 and 20 mHz at the chromosphere, and, thus, the contribution of waves to chromospheric heating in sunspot could be lower. In our simulations we have rejected the waves with frequencies above 20 mHz. According to \citet{Fossum+Carlsson2005} and \citet{Carlsson+etal2007} the power of waves with 20 mHz frequency can be assumed to be negligible for chromospheric heating, but \citet{WedemeyerBohm+etal2007} and \citet{Kalkofen2007} have pointed out that this question depends on the spatial resolution of the data used.

Previous works have found that the acoustic wave energy is too low, by a factor of at least ten, to balance the radiative losses in the non-magnetic solar chromosphere \citep{Fossum+Carlsson2005,Carlsson+etal2007}. In the context of small-scale magnetic flux concentrations, \citet{Vigeesh+etal2009} obtained that the acoustic energy flux of transversely, impulsively excited waves is insufficient to balance the chromospheric radiative losses in the network. The fraction of the required energy supplied by acoustic waves in the magnetized atmosphere of a sunspot seems to be larger than in the quiet Sun. The chromospheric energy losses in umbral regions are 3-5 times lower than in quiet Sun, but the energy supplied by the waves seems to be similar. At the photosphere, the surface amplitude of solar p-modes in magnetic regions are reduced below those in magnetically quiet regions in the 5 minute band \citep{Woods+Cram1981}. \citet{Brown+etal1992} showed that the supression is frequency dependent, peaking at 4 mHz. The reduction of the amplitude occurs for frequencies below 5 mHz, while the amplitude of waves above 5.5 mHz is enhanced. Since the waves that propagate energy to higher layers are those above the cutoff frequency, the reduction of p-modes in regions of magnetic activity does not produce a lower efficiency of acoustic energy propagation in sunspots. Several works have pointed out that the fast magnetic mode in the low-$\beta$ region is reflected back toward the photosphere and it is unable to supply energy to higher layers \citep{Rosenthal+etal2002,Khomenko+Collados2006, Felipe+etal2010a}. According to \citet{Cally2005}, the fast-to-fast conversion is more efficient for higher frequencies and, thus, waves with frequencies above the cutoff are significantly transformed into fast magnetic waves above the height where $v_A=c_S$ and cannot carry energy to the chromosphere. In our simulation the energy flux of the fast magneto-acoustic mode at the chromosphere amounts to $2\times10^3$ erg cm$^{-2}$ s$^{-1}$, being 2-3 orders of magnitude lower than the acoustic flux.  The LOS velocity oscillations measured with the \SiI\ line correspond to the slow acoustic waves in the \hbox{low-$\beta$} region, after the transformation of the upward propagating waves. Their power has already been reduced by the transformation to the fast magnetic mode. \hbox{Figure \ref{fig:Fac_height}} shows that at the formation height of the \SiI\ line, the average acoustic energy flux is around \hbox{10$^{6}$ erg cm$^{-2}$ s$^{-1}$}, so at this photospheric height the energy contained in the form of acoustic flux is of the order of magnitude of the amount required by the chromospheric radiative losses.

As a further step in our work, we aim to synthetize the spectral lines from the numerical simulation in order to develop a better comparison with the spectropolarimetric observations. This process is relatively simple in the case of the photospheric lines, since they are formed at LTE, but the non-LTE formation of the chromospheric lines hinders their synthesis. The last case is especially interesting to clarify the issue of the variation in the velocity response height of the atmosphere to nonlinear perturbations. On the other hand, we also plan to perform similar simulations using a bidimensional velocity field obtained with IBIS or HMI/SDO as a driver.

\acknowledgements   This research has been funded by the Spanish
MICINN through projects AYA2007-66502 and AYA2010--18029. The
simulations have been done on LaPalma supercomputer at Centro de
Astrof\'{\i}sica de La Palma and on MareNostrum supercomputer at
Barcelona Supercomputing Center (the nodes of Spanish National
Supercomputing Center)

\aareferences

\end{document}